\documentclass[aps,pre,twocolumn,superscriptaddress]{revtex4-1}
\usepackage[dvipdfm]{graphicx}
\usepackage{amssymb,amsmath}
\usepackage{bm}

\begin{document}


\title{Potential global jamming transition in aviation networks}



\author{Takahiro Ezaki}
\email{ezaki@jamology.rcast.u-tokyo.ac.jp}
\affiliation{Department of Aeronautics and Astronautics, Graduate School of Engineering,
The University of Tokyo, 7-3-1 Hongo, Bunkyo-ku, Tokyo 113-8656, Japan}
\affiliation{Japan Society for the Promotion of Science, 8 Ichibancho, Kojimachi, Chiyoda-ku, Tokyo 102-8472, Japan}

\author{Katsuhiro Nishinari}
\affiliation{Research Center for Advanced Science and Technology,
The University of Tokyo, 4-6-1, Komaba, Meguro-ku, Tokyo 153-8904, Japan}


\date{\today}

\begin{abstract}
In this paper, we propose a nonlinear transport model for an aviation network.
The takeoff rate from an airport is characterized by the degree of ground congestion. 
Due to the effect of \textit{surface congestion}, the performance of an airport 
deteriorates because of inefficient configurations of waiting aircraft on the ground. 
Using a simple transport model, we performed simulations on a U. S. airport network
and found a global jamming transition induced by local surface congestion. 
From a physical perspective, the mechanism of the transition is studied analytically 
and the resulting aircraft distribution is discussed considering system dynamics. 
This study shows that the knowledge of the relationship between a takeoff rate and a congestion level on the ground is vital for efficient air traffic operations.
\end{abstract}

\pacs{}

\maketitle


\section{Introduction}
\begin{figure}[htbp]
 \begin{center}
  \includegraphics[width=80mm]{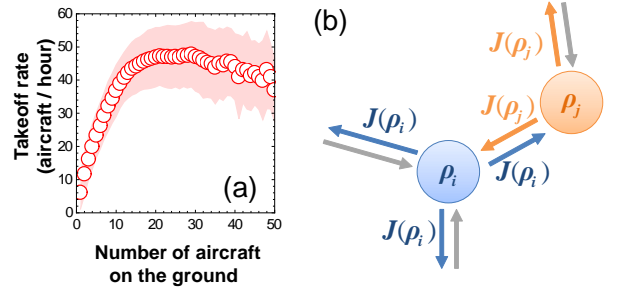}
 \end{center}
 \caption{(a) Empirical relationship between the takeoff rate and surface congestion at Philadelphia International Airport (data obtained from Ref. \cite{Simaiakis}.). The shaded area indicates the standard deviation of the data. (b) Flow between two nodes in the network model.}
 \label{rule}
\end{figure}

Faced with rapidly growing demand in air traffic, researchers have addressed 
aggravating congestion problems from various perspectives \cite{Filar, Barnhart, Neuf, Ahmad, Clausen, Lordan1, Petersen}.
Airport congestion related to departing aircraft on the ground, i.e., \textit{surface congestion}, is known to be a serious problem, and its relationship to airport performance has been recently investigated \cite{Idris, Simaiakis}. 
Figure \ref{rule}(a) shows the empirical takeoff rate at Philadelphia International Airport presented by Simaiakis and Balakrishnan \cite{Simaiakis}. When the airport is not congested, increasing the number of aircraft on the ground directly leads to a high takeoff rate. However, after the number reaches a critical point, the performance deteriorates due to
inefficient aircraft configurations. Note that such unimodal relationships can be universally observed in the motion of self-propelled particles \cite{Chowdhury,Helbing} (e.g., vehicular traffic \cite{Hall, Greenshields}, pedestrian flow \cite{Weidmann,Armin}, and interacting particle systems \cite{Derrida, Schadschneider}).

In this paper, we aim to disclose the physical aspects of air traffic from the surface congestion viewpoint.
For this purpose, we consider a network of airports (nodes), the performance of which is determined by their congestion levels, as shown in Fig. \ref{rule}(b). 
Note that network approaches to describe air traffic have been proposed in previous studies \cite{Bertsimas,Menon,Sridhar}; however, these studies did not focus on the surface congestion effect.  
In addition to the conventional approaches in the field of management science, 
the network modeling of aviation transport has recently been attempted \cite{Lacasa, Fleurquin}
from a network science \cite{Strogatz, Albert, Newman, Boccaletti} perspective,
supported by the recent understanding of airport networks \cite{Li, Guimera1, Guimera2, Zanin, Bagler}. However, the knowledge of congestion dynamics from a network perspective is still limited. 

This study presents a simple model for aviation transport networks based on the density--flux relationship locally defined in each airport. Our primary goal is to evaluate the stability of 
global air traffic against disturbances. 
Because of assumptions imposed in the model, we cannot expect a quantitative description of the real systems. 
However, taking advantage of its simplicity, we aim to clarify the fundamental features of a spontaneous jam formation.

The remainder of this paper is organized as follows. The next section defines the model and simulation conditions. In Section \ref{result}, we show a spontaneous jamming transition phenomenon and its dynamics. 
Finally, in Section \ref{disc}, we discuss the implication of these phenomena and future work.

\section{Model}\label{model}
\subsection{Problem formulation}
Consider $N$ nodes, labeled $1,\cdots,N$, representing each airport. 
The connectivity of the nodes is defined by an adjacency matrix $\bm{A}$, 
whose element $A_{ij}$ indicates the presence ($A_{ij}=1$) or absence ($A_{ij}=0$) of a link between 
the two nodes $i$ and $j$ ($i,j=1,\cdots,N$). 
Time development of the number of aircraft in node $i$, $n_i$, is described by the equation of continuity as follows:
\begin{eqnarray}
\frac{\partial n_i}{\partial t} &=& \sum_{j=1}^{N}{A_{ji} J(\rho_j)} -\sum_{j=1}^{N}{A_{ij} J(\rho_i)},
\end{eqnarray}
where $\rho_i$ and $J$ denote the density in node $i$ and outflux from the node, respectively.
Here, for the sake of simplicity, we assumed that each link has an identical capacity.
By assuming that each node has a capacity that is proportional to the degree, $k_i=\sum_{j=1}^N{A_{ij}}$,
and defining $\rho_i = n_i/k_i$ \cite{densitydef},
the time evolution of a density set, $\{\rho_1,\cdots\,\rho_N\}$, is written as
\begin{eqnarray}
\frac{\partial \rho_i}{\partial t} &=& \frac{1}{k_i}\sum_{j=1}^{N}{A_{ji} J(\rho_j)} - J(\rho_i).\label{td}
\end{eqnarray}
In the context of vehicular traffic, this formulation is known as the \textit{cell transmission model} proposed by Daganzo \cite{Daganzo1, Daganzo2,Daganzo3},
in which each road segment defines a node.
Previous studies focused on a simple graph comprising a small number of nodes, and very little is known about the dynamics on complex networks.
Note that Sun and Bayen \cite{Sun} applied this cell transmission model to air traffic in a different manner, in which nodes correspond to discretized aviation routes.


\subsection{Simulation conditions}
We use $A_{ij}$ defined from the U. S. airport network \cite{Colizza}. 
The data comprise 500 major airports and 2980 undirected links ($A_{ij}=A_{ji}$), 
and the network is connected (i.e., there is at least a single path of links between any two different nodes).
In this paper, we define an explicit form of flux--density relation as follows:
\begin{equation}
J(\rho_i)= \begin{cases}
\rho_i(1-\rho_i)  & \text{if $\rho_i< \rho_{\rm{max}}$,}\\
\rho_{\rm{max}}(1-\rho_{\rm{max}}) & \text{if $\rho_i\geq\rho_{\rm{max}}$,}
\end{cases}\label{fdf}
\end{equation}
where $\rho_{\rm{max}}$ is the maximum density that a single airport can accommodate on the ground. Above this density, aircraft have to stay in the sky near the airport (they still belong to the destination node); $J$ does not decrease since the congestion level on the ground remains the same.
Note that in Fig. \ref{rule}(a), only aircraft on the ground are counted (i.e., $\rho_{\rm{max}}$ corresponds to the right end of the figure.). 
We assumed a quadratic function to describe this part as the simplest smooth unimodal function.
This function is often used to model dynamics in which capacity constraints play a role (e.g., vehicular traffic \cite{Greenshields}, interacting particle systems \cite{Schadschneider}, and population dynamics \cite{May}).

This function is illustrated in Fig. \ref{fds}(a), in which we set $\rho_{\rm{max}}=0.7.$
We define the critical density, $\rho_{\rm{cr}}$, as the density that gives maximum $J$.
In the function (\ref{fdf}), $\rho_{\rm{cr}} = 0.5.$

At $t=0$, the initial density of nodes are set almost uniformly as $\bar{\rho}$, which is slightly fluctuated with a random variable $\delta$ drawn from a uniform distribution $[-0.005,0.005]$. Each simulation is performed
until the time development of densities stops. 
We study the characteristics of the stationary state by investigating key quantities, which will be defined subsequently.
Unless otherwise noted, results are averaged over $1 000$ trials.

\section{Results}\label{result}
\subsection{Global flux-density relationship}
We define the global flux, $\bar{J}$, as the average flow per link as follows:
\begin{equation}
\bar{J}(\bar{\rho}) = \frac{\sum_{i=1}^N\sum_{j=1}^N{A_{ij}J(\rho_i^\infty)} }{\sum_{i=1}^N{k_i}},
\end{equation}
where $\rho_i^\infty$ denotes the density $\rho_i$ in a stationary state.
When the density is uniformly distributed in the stationary state, $\bar{J}(\bar{\rho})$ coincides with the local flow $J(\bar{\rho})$ at $t=0.$
However, as illustrated in Fig. \ref{fds}(a), one can find a gap between $\bar{J}(\bar{\rho})$ and $J(\bar{\rho})$ for a certain range of $\bar{\rho}$ due to the nonuniform distribution of $\rho_i^\infty$. 
For $\bar{\rho}<\rho_{\rm{cr}}=0.5$ and $\bar{\rho}>\rho_{\rm{max}}=0.7,$ the uniformity of $\rho_i^{\infty}=\bar{\rho}$ is conserved for all nodes, whereas for the middle initial density, 
some nodes take a single constant value of $\rho_i=\rho_{l}$ while the others distribute in the range 
of $\rho_i \geq \rho_{\rm{max}}$ [Fig. \ref{fds}(b)]. 
The lower bound density, $\rho_l$, is the density value that satisfies $J(\rho_l)=J(\rho_{\rm{max}})\; (\rho_l<\rho_{\rm{cr}})$ [see Fig. \ref{sta}(b)]. 
Interestingly, at the critical density ($\bar{\rho}=0.5$), only a small number of nodes draw density from neighboring nodes.

This jamming transition can be understood as a bifurcation phenomenon.
\subsection{Global jamming transition}
Here, we investigate the stability of the system state that has uniform density distribution, $\rho_i=\bar{\rho}\quad(i=1,\cdots,N)$, by considering the behavior of a small perturbation $\epsilon_{i}$ [we also define a vector, $\bm{\epsilon}=(\epsilon_1,\cdots ,\epsilon_N)^T$] added to the state as $\rho_i=\bar{\rho}+\epsilon_i$.
Linearizing Eq. (\ref{td}) yields  
\begin{eqnarray}
\frac{\partial \epsilon_i}{\partial t} &=& \frac{1}{k_i} \sum_{j=1}^{N}{J'(\bar{\rho})A_{ji}\epsilon_j}- J'(\bar{\rho})\epsilon_i,\\
\frac{\partial}{\partial t} \bm{\epsilon} &=&- J'(\bar{\rho})\bm{M\epsilon},
\end{eqnarray}
where matrix $\bm{M}$ is defined by $M_{ij} = A_{ji}/k_i\;(i\neq j)$ and $M_{ii}=-1$.
The stability of the system is directly connected to the absence of the positive eigenvalues of the matrix, $-J'(\bar{\rho})\bm{M}$. By applying the Perron--Frobenius theorem to $\bm{M+I}$ ($\bm{I}$ is a unit matrix), 
one can find that its eigenvalues $\lambda_i\;(i=1,\cdots,N)$ are not larger than $1$. 
The eigenvalues of the matrix $M$ are given by $\lambda - 1$. Note that by definition, $\bm{A},\bm{M},$ and $\bm{M+I}$ are all irreducible, and each element of $\bm{M+I}$ is not less than zero; thus, the Perron-Frobenius theorem can be applied.
Therefore, the stability is dependent only on $J'(\bar{\rho})$, i.e., the system is unstable when  $J'(\bar{\rho})<0$.
Thus, \textit{local} jamming in the fundamental relationship $J(\bar{\rho})$ directly triggers a \textit{global} jam. 

\begin{figure}[tbp]
 \begin{center}
  \includegraphics[width=70mm]{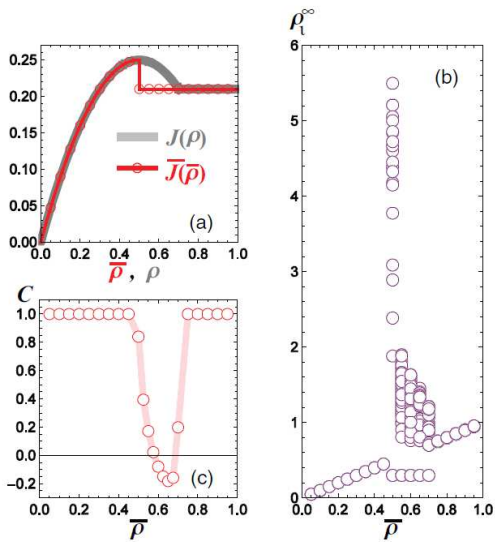}
 \end{center}
 \caption{(a) Local ($J(\rho)$; gray) and global ($\bar{J}(\bar{\rho})$; red) fundamental diagrams. (b) Stationary density distribution of a single trial. Each circle represents the density in a single node.
(c) Global correlation coefficient. When the initial density $\bar{\rho}$ is small and large, all nodes have the same density value, which is reflected in the agreement of the local and global fundamental diagrams [panel (a)].
On the other hand, when $\bar{\rho}$ is in the middle regime ($\rho_l<\bar{\rho}<\rho_{\rm{max}}$), the uniformity of the density is broken [panel (b) and (c)], and the global flux is deteriorated [panel (a)].
} \label{fds}
\end{figure}

\begin{figure}[tbp]
 \begin{center}
  \includegraphics[width=80mm]{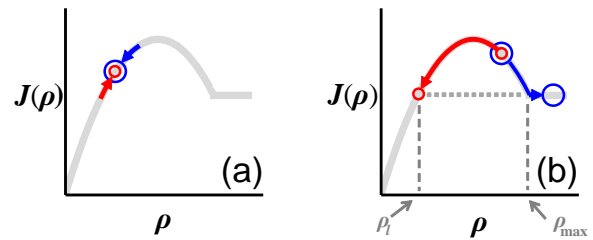}
 \end{center}
 \caption{Stability of  (a) subcritical and (b) supercritical density.}
 \label{sta}
\end{figure}

The flux deterioration shown in Fig. \ref{fds}(a) can be understood as follows. When the system is not congested [Fig. \ref{sta}(a)], $\rho_i = \bar{\rho}$ is locally stable and the density is uniformly distributed to all nodes. 
Conversely, in high density cases [Fig. \ref{sta}(b)] this state is no longer stable, 
and the flow between nodes equilibrates at a lower density value $\rho_i=\rho_l$ and higher density value $\rho_i\geq\rho_{\rm{max}}.$
Hereafter, we refer to this state as the \textit{globally jamming state}.

\subsection{Global density correlation}
In the previous subsection, we found that the density distribution is imbalanced.
To evaluate the regularity in this stationary density distribution, we define an indicator variable $\sigma_{i}$,
which reflects the density in node $i$:
\begin{equation}
\sigma_{i} = 
\begin{cases}
1  & \text{if $\rho_i\geq \rho_{\rm{cr}}$,}\\
-1 & \text{if $\rho_i<\rho_{\rm{cr}}$.}
\end{cases}
\end{equation}
Using this variable, correlation between two neighboring nodes is quantified as $A_{ij}\sigma_{i}\sigma_{j},$
which returns $1$ if two neighboring nodes have the same $\sigma$ and returns $-1$ otherwise. 
Figure \ref{fds}(c) illustrates global regularity of the density distribution defined as the average correlation degree per link
\begin{equation}
C = \left\langle\frac{\sum_{i=1}^N\sum_{j=1}^N{A_{ij}\sigma_i\sigma_j}}{\sum_{i=1}^N k_i}\right\rangle,
\end{equation}
where $\langle x\rangle$ denotes an ensemble average of $x.$
Corresponding to the aforementioned bifurcation, $C$ is constant ($C=1$) for $\bar{\rho}<\rho_{\rm{cr}}$ and $\bar{\rho}>\rho_{\rm{max}}$,
and in the middle density regime, it suddenly decreases to approximately $0$, explaining the breakage of the uniform state. 
From this value we can conclude that the global regular structure is absent in terms of density correlation in the globally jamming state. 

\subsection{Local density correlation and dynamics}
\begin{figure}[tbp]
 \begin{center}
  \includegraphics[width=70mm]{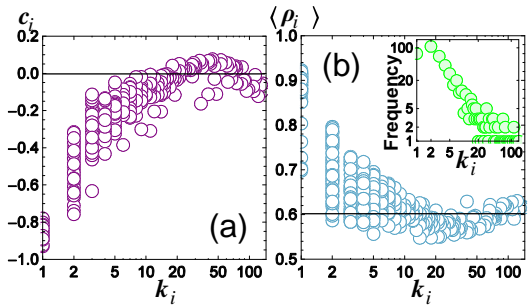}
 \end{center}
 \caption{(a) Local correlation coefficient. (b) Average density distribution. Each plot corresponds to averaged stationary density
in one node. The inset in panel (b) shows degree distribution of the focal network.}
 \label{ciri}
\end{figure}

Next, we focus on the local regularity of the stationary density distribution. Here, we set $\bar{\rho} = 0.6$ and 
investigate the average local correlation per node
\begin{equation}
c_{i} = \left\langle \frac{\sum_{j=1}^{j=N}A_{ij} \sigma_i\sigma_j }{k_{i}}\right\rangle.
\end{equation}
As shown in Fig. \ref{ciri}(a), $c_i$ is highly dependent on $k_i$. 
Decreasing values of $k$ correspond to values of  $c_i$ decreasing to near $-1$, which indicates that
 a node tends to have an opposite $\sigma$ against its neighboring nodes with high probability.
However, as the number of neighboring nodes increases, it becomes difficult to take the opposite value
against all the neighboring nodes, and thus the local correlation vanishes.
In other words, the presence of high-degree nodes reduces the absolute value of the local and global correlations in the globally jamming state.

\begin{figure}[tbp]
 \begin{center}
  \includegraphics[width=70mm]{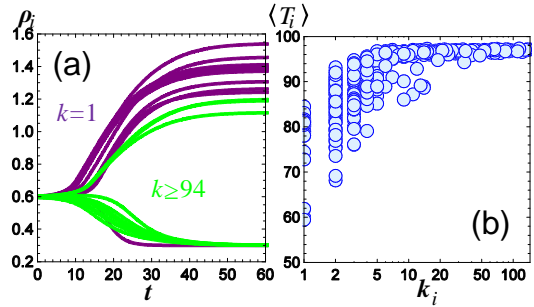}
 \end{center}
 \caption{(a) Sample time series of density for large ($k\geq 94$) and small ($k=1$) nodes. 
(b) Average relaxation time of each node. Each plot corresponds to averaged relaxation time of one node.
}
 \label{ts}
\end{figure}

In addition, the density also depends on $k_i$ [Fig. \ref{ciri}(b)]. 
In hub nodes (large nodes), the average density is approximately $0.6$, which coincides with $\bar{\rho}$, 
whereas in nodes with small $k_i$ (small nodes), the average density is significantly larger.
This can be explained by the effective relaxation speed of the density [Figs. \ref{ts}(a) and (b)]. 
This dependency is supported by Fig. \ref{ts}(b), which shows the average of saturation time $T_i = \min{\{t | \rho_i(t+1) = \rho_i(t)\}}.$
This is because, for a large node, the density inflow to the node is averaged over a large number 
of neighboring nodes and its speed of variation is relatively slow. 
Therefore, at the beginning of each simulation trial in the middle density regime, the change in density is dominant in small nodes,
which is strongly influenced by the initial perturbation $\delta$: for positive $\delta$, the density starts to increase, and for negative $\delta$ density starts to decrease. Thus, one half of the small nodes become congested and the other half loses density. Since the drag of density is faster than discharge [Fig. \ref{ts}(a)], and there is no upper limit on the density, the average final density in small nodes becomes large. 
Conversely, in large nodes, at the beginning, it tends to be deprived of density neighboring small nodes, which leads to a less probability of congestion occurring in these nodes.
As a result, the density in large nodes is suppressed.
 
\section{Discussions}\label{disc}
We proposed a transportation model for aviation net, focusing on the surface congestion effect.
The model exhibits a jamming transition, which is connected to a bifurcation of the system equations. Interestingly, when each component of the network (airport) is jammed, characterized as a negative differential coefficient of a flux-density relationship, the uniform density distribution becomes unstable, decreasing the total network flux (global jamming). Through relaxation from a uniform to an imbalanced distribution, densities on two neighboring nodes tend to vary inversely, i.e., one increases and the other decreases. 
However, on a node having large number of links, this correlation vanishes, breaking global correlation structures. 
In this model, density in a large node changes more slowly than that in a small node because the density inflow to the node is averaged over many nodes
 and does not change sharply. 
Note that because the model assumes that the capacity of each node is proportional to its degree, 
the time constant of the density variation is normalized against the degree.
Because of the difference in time coefficient, small nodes are more congested on average.
In a small node, however, there may be small traffic demand, which contradicts the assumption that each link has the same amount of traffic. Hence if we take the traffic capacity of links into consideration,
the time constant in small nodes is reduced and thus the final density distribution becomes more uniform.   

On the other hand, the global jamming transition presented in this study is not influenced by 
these simplifications because it is caused by a fundamental bifurcation observed even in a two-component system \cite{Daganzo3}. 
In previous studies, this type of jamming transition in air traffic systems has rarely received attention.
An effective method to mitigate this jamming is not to accept airplanes beyond the point at which the airport performance decreases. This can be realized by rerouting techniques and the Ground Delay Program (GDP) \cite{GDP} operated by the Federal Aviation Administration \cite{Luo,Rosenberger,Lan}.   
Therefore, knowing the flux--density relationship in each airport for operational reference is vital.

Note that the jamming presented in this study is a completely different phenomenon from the delay propagation phenomena that have been actively studied \cite{Fleurquin, Ahmad, Luo,Rosenberger,Lan}. In these studies, a main cause of delay is the crew and passenger connection disruptions. This type of delay propagates downstream. In this study, congestion in one airport does not simply propagate; it disrupts the density in neighboring airports through its outflow traffic. This study also provides a reason why a small disturbance, for example, due to a bad weather, grows to
a large delay from a systemic perspective. Furthermore, note that if a congested airport stops accepting
aircraft by following a strategy such as the GDP, congestion propagates to upstream airports. 
Thus, the propagation direction is determined by its cause. 

Our results are based on a model that only considers fixed physical constraints (surface congestion curve) and ignores flight schedules and operational policies.  In the future, we intend to address congestion problems on aviation networks more comprehensively by integrating all these factors,
which we believe will provide insight to the problem.

We wish to thank Naoki Masuda for discussions on this study. 
We would also like to acknowledge Daichi Yanagisawa and Ryosuke Nishi for their valuable comments on this paper. TE received support from the Japan Society for the Promotion of Science, Grants-in-Aid for Scientific Research (Grant No. 13J05086).

\bibliography{basename of .bib file}

\begin{thebibliography}{00}
\bibitem{densitydef}Note that here the ``density'' is standardized to reduce the number of parameters as follows. 
Using the assumption that the capacity of node $i$, $n_{i,\rm{max}}$, is proportional to $k_i$, i. e., 
$n_{i,\rm{max}} = ak_i$ ($a$: const.), the occupation degree, $\tilde{\rho}_i,$ can be characterized as 
$\tilde{\rho}_i = \frac{n_i}{ak_i}.$ We study the system based on $\rho_i = a\tilde{\rho}_i$, instead of $\tilde{\rho}_i.$ The real aircraft number corresponding to $\rho_i$ can be calculated from the surface congestion curve of the airport.
\bibitem{GDP}A program that controls departure time of aircrafts on the ground when a destination airport cannot accept them because of bad weather, heavy congestion or an aircraft accident.
\end{thebibliography}

\end{document}